\documentclass[aps,prl,twocolumn,showpacs,superscriptaddress]{revtex4}

\usepackage{amsmath}
\usepackage{graphicx,color}

\newcommand{\etal}{{\textit{et al.}}}

\newcommand{\pt}{p_T}
\newcommand{\nn}{\nonumber\\}
\newcommand{\ba}{\begin{eqnarray}}
\newcommand{\ea}{\end{eqnarray}}
\newcommand{\la}[1]{\label{#1}}

\newcommand{\gm}{\gamma}

\begin{document}


\title{Medium Modification of $\gamma$ Jets in High-Energy Heavy-Ion Collisions }


\author{Xin-Nian Wang}
\affiliation{Key Laboratory of Quark and Lepton Physics (MOE) and Institute of Particle Physics, Central China Normal University, Wuhan 430079, China}
\affiliation{Nuclear Science Division Mailstop 70R0319,  Lawrence Berkeley National Laboratory, Berkeley, California 94740, USA}
\author{Yan Zhu}
\affiliation{Key Laboratory of Quark and Lepton Physics (MOE) and Institute of Particle Physics, Central China Normal University, Wuhan 430079, China}
\affiliation{Nuclear Science Division Mailstop 70R0319,  Lawrence Berkeley National Laboratory, Berkeley, California 94740, USA}
\affiliation{Faculty of Physics, University of Bielefeld, D-33501 Bielefeld, Germany}


\date{\today}

\begin{abstract}

 Two puzzling features in the experimental study of jet quenching in central Pb+Pb collisions at the LHC
 are explained within a linearized Boltzmann transport model for jet propagation.
 A $\gamma$-tagged jet is found to lose about 15\% of its initial energy while its azimuthal angle remains almost unchanged  
 due to rapid cooling of the medium. The reconstructed jet fragmentation function is found to have some modest enhancement at both small and large fractional momenta as compared to that in the vacuum because of the increased contribution of leading particles to 
the reconstructed jet energy and induced gluon radiation and recoiled partons. A $\gamma$-tagged jet fragmentation function is proposed that is more sensitive to jet-medium interaction and the jet transport parameter in the medium. The effects of
recoiled medium partons on the  reconstructed jets are also discussed.

\end{abstract}

\pacs{25.75.Bh,25.75.Cj,25.75.Ld}

\maketitle

Parton energy loss due to multiple scattering and bremsstrahlung in dense medium
should be accompanied by transverse momentum ($p_T$) broadening \cite{Baier:1996sk}.
One therefore should expect to see suppression of the yield and $p_T$ broadening of leading hadrons from jet fragmentation. This is the main mechanism behind the observed jet quenching  \cite{Wang:1991xy} phenomena as manifested in the suppression of large $p_{T}$ hadron spectra, dihadron and $\gamma$-hadron correlations in high-energy
heavy-ion collisions. 

The jet quenching study has also been extended to full jets \cite{Vitev:2009rd}, which are reconstructed with a jet-finding algorithm \cite{Cacciari:2011ma} and consist of collimated clusters of hadrons (or partons in a partonic description) within a jet cone $\sqrt{(\phi-\phi_J)^2+(\eta-\eta_{J})^2}\leq R$, where $\eta$ ($\eta_J$)  and $\phi$ ($\phi_J$) are hadrons'  (jet's) pseudorapidity and azimuthal angle, respectively.  Since some jet shower partons can be transported outside the jet cone through multiple scattering and bremsstrahlung, one should also expect to see a reduction of the reconstructed jet energy and change of its azimuthal angle. In addition, one also expects to see a modification of the jet fragmentation function and jet transverse profile.
A large dijet asymmetry in central Pb+Pb collisions at $\sqrt{s}=2.76$ TeV is indeed observed at the Large Hadron Collider (LHC) \cite{Aad:2010bu}, consistent with the picture of jet quenching \cite{Qin:2010mn,Young:2011qx,He:2011pd,Renk:2012cx,Zapp:2012ak}. However, two puzzles in jet modification still lack satisfactory explanations.  Despite the large dijet asymmetry, there is no apparent azimuthal angle broadening within experimental errors. The reconstructed jet fragmentation function has a modest but interesting modification with enhancement at both large and small momentum fractions  $z_{\rm jet}=p_{L}/E_{\rm jet}$ \cite{CMS:2012wxa}.

Back-to-back $\gamma$ jets are considered ``golden channels'' for the study of jet quenching since they have less trigger bias \cite{Wang:1996yh} than dijets. Although $\gamma$-jet asymmetry as measured in Pb+Pb collisions at the LHC \cite{Chatrchyan:2012gt, ATLAS:2012cna} can be explained by parton energy loss \cite{Dai:2012am,Qin:2012gp}, there is still a lack of satisfactory explanations of the apparent puzzles in the structure of the medium modified jets.
In this Letter, we will report a first study of medium modification of $\gamma$-tagged jets  within a linearized Boltzmann transport (LBT) model \cite{Li:2010ts}  and address the aforementioned puzzles in the measured jet structure in high-energy heavy-ion collisions at the LHC. 
We further propose measurements of the $\gamma$-tagged jet fragmentation function and its medium modification
that are more sensitive to jet-medium interaction and the jet transport parameter.  
We will also discuss the effects of jet-induced medium excitation since the LBT model tracks the transport of both shower and recoiled partons.

Within the LBT model, the propagation of jet shower partons and medium excitation is simulated according to a linearized
Boltzmann equation
\ba
p_1\cdot\partial f_1(p_1)&=&-\int dp_2dp_3dp_4 (f_1f_2-f_3f_4)|M_{12\rightarrow34}|^2
\nn &\times&
(2\pi)^4\delta^4(p_1+p_2-p_3-p_4),
\ea
where $dp_i=d^3p_i/[2E_i(2\pi)^3]$, $f_i=1/(e^{p\cdot u/T}\pm1)$ $(i=2,4)$ are parton phase-space distributions in a thermal medium with local temperature $T$ and fluid velocity $u=(1, \vec{v})/\sqrt{1-\vec{v}^2}$, and $f_i=(2\pi)^3\delta^3(\vec{p}-\vec{p_i})\delta^3(\vec{x}-\vec{x_i}-\vec{v_i}t)$ $(i=1,3)$ are the parton phase-space densities before and after scattering.
We assume a small angle approximation for the elastic scattering amplitude $|M_{12\rightarrow34}|^2=C g^4(\hat s^2+\hat u^2)/(-\hat t+\mu^2_{D})^2$ with Debye screening mass $\mu_{D}$, where $\hat s$, $\hat t$, and $\hat u$ are Mandelstam variables, and $C=1$ (9/4) is the color factor for quark-gluon (gluon-gluon) scattering.  The strong coupling constant $\alpha_{ s}=g^{2}/4\pi$ is fixed and will be determined via comparisons to experimental data.

Partons are assumed to propagate along classical trajectories between two adjacent collisions. The probability of scattering is determined in each time step $\Delta t$ by $P_{a}=1-\text{exp}[-\sum_j (\Delta x_j\cdot u)\sum_{b}\sigma_{ab}\rho_{b}(x_j)]$, where $\sigma_{ab}$ is the
parton scattering cross section, the sum over time steps starts from the last scattering point, and $\rho_{b}$ is the local medium parton density. Both shower $(p_3)$ and recoiled medium partons $(p_4)$ after each scattering are followed by further scatterings in the medium. To account for the backreaction in the Boltzmann transport, initial thermal partons $(p_2)$, denoted as ``negative'' partons, are transported according to the Boltzmann equation. Their energies and momenta will be subtracted from all final observables. These negative partons are considered as part of the recoiled partons that are responsible for jet-induced medium excitations \cite{Li:2010ts}. 

In this study, the LBT model is extended to include induced radiation accompanying each elastic scattering according to the high-twist approach \cite{Wang:2001ifa}
\ba \la{induced}
\frac{dN_g^{a}}{dzdk_\perp^2dt}=\frac{6\alpha_sP(z)}{\pi k_\perp^4}(\hat p\cdot u) \hat{q}_{a} \sin^2\frac{t-t_i}{2\tau_f},
\ea
where $z$  and $k_\perp$ are the energy fraction and transverse momentum of the radiated gluon, $\hat p_{\mu}=p_{\mu}/p_{0}$, $P(z)=[1+(1-z)^2]/z$ the splitting function,  $\tau_f=2Ez(1-z)/k_\perp^2$ the gluon formation time, and $\hat{q}_{a}=\sum_{b}\rho_{b}\int d\hat t q_\perp^2 d\sigma_{ab}/d\hat t $ the jet transport parameter. The Debye screening mass $\mu_D$ is used as an infrared cutoff for the gluon's energy. Multiple gluon emissions induced by a single  scattering are included according to a Poisson distribution.
All radiated gluons are assumed to be on-shell, and their 4-momenta are successively 
determined from Eq.~(\ref{induced}). 


For initial configurations of $\gamma$ jets, we use {\footnotesize HIJING} \cite{Wang:1991hta} for $p+p$ collisions at $\sqrt{s}=2.76$ TeV with a trigger on the transverse momentum transfer $q_{T}\ge 30$ GeV in the c.m. frame of two colliding partons. Further selections are made for events with $p_{T}^\gamma > 60 $ GeV. Jet shower partons,  including those from both initial- and final-state radiation, are transported through a thermal medium within the LBT model. The final partons, including negative partons, are used for jet reconstruction
using a modified version of the anti-$k_t$ algorithm in {\footnotesize FASTJET} \cite{Cacciari:2011ma}, in which energies and momenta of
negative partons are subtracted from the final jet observables. The energies of jets reconstructed from the final hadrons and partons differ only about 1 GeV in $p+p$ collisions from {\footnotesize HIJING}.

According to Eq.~(\ref{induced}), the radiative energy loss of a single parton going through multiple scattering is
\begin{equation}
\Delta E_a\approx \frac{3\alpha_{ s}}{2} \int d\tau (\tau-\tau_0) (\hat p\cdot u) \hat{q}_{a} \ln \frac{2E}{(\tau-\tau_0)\mu_D^2}.
\end{equation}
The corresponding $p_T$ broadening is
\begin{equation}
\langle \Delta p_T^2\rangle = \int d\tau (\hat p\cdot u) \hat{q}_{a} .
\end{equation}
In a static and uniform medium, the total parton energy loss has an approximate quadratic dependence on the propagation length while the $p_T$  broadening has a linear dependence. In high-energy heavy-ion collisions, the jet transport parameter should have a time dependence $\hat q_a = \hat q_a^0 (\tau_0/\tau)^{1+\alpha}$ with $\alpha\ge 0$ whose value becomes bigger in the later stage of evolution due to fast 3D expansion. In this case, the total energy loss is approximately linear in the propagation length while the $p_T$ broadening has a logarithmic dependence or less.  These length dependences should be the same for reconstructed jets as we will show. It is important to keep in mind that fast expansion in heavy-ion collisions results in much bigger reduction in the
$p_T$ broadening than the jet energy loss.

\begin{figure}
\centerline{\includegraphics[width=8.9cm]{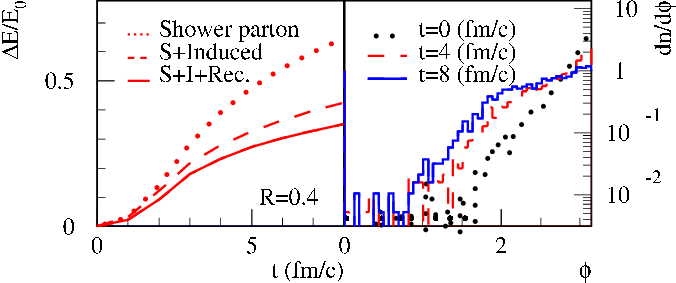}}
 \caption{ Energy loss as a function of time (left) and
 azimuthal distribution relative to the $\gamma$ direction (right) for $\gamma$-tagged jets in a 
 uniform gluonic medium.
 \label{uniform}}
\end{figure}

To illustrate the connection between energy loss and azimuthal broadening for reconstructed jets, we consider first the propagation of $\gamma$-tagged jets in a uniform and static gluonic medium at temperature $T=300$ MeV. We set $\alpha_s=0.4$ and $\mu^2_{D}=1$ GeV$^2$ for simplicity and use a jet-cone size $R=0.4$.  Shown in Fig.~\ref{uniform} (left) is the jet energy loss that has a clear quadratic time dependence during the early times. The corresponding $p_T$ broadening of single partons should increase linearly during early times. This leads to a significant azimuthal
angle broadening of the reconstructed jets, as shown in Fig.~\ref{uniform} (right).
If we include only jet shower partons (dotted line) in the jet reconstruction, the energy loss is considerably larger than when radiated gluons (dashed line) or all (shower + radiated + recoiled ) partons (solid line) are included. Both radiated gluons and recoiled medium partons enter the jet cone and become part of the reconstructed jets. 

For the study of $\gamma$-tagged jets in heavy-ion collisions within the LBT model,  
we use the space-time profile of temperature and fluid velocity in the quark-gluon phase from (3+1)D ideal hydrodynamical 
simulations \cite{Hirano:2005xf} of  Pb+Pb collisions at the LHC. 
The initial $\gamma$-jet production from {\footnotesize HIJING} is distributed according to the overlap function of two colliding nuclei with a Wood-Saxon nuclear geometry. Whenever comparisons are made to the experimental data, we apply the same
kinematic cuts to the LBT results. For CMS data \cite{Chatrchyan:2012gt},  $\pt^{\gamma}>60$ GeV, $|\eta^{\gamma}|<1.44$, $\pt^\text{jet}>30$ GeV, $|\eta^\text{jet}|<1.6$, and $\Delta\phi=|\phi^\text{jet}-\phi^\gm|>7\pi/8$, and for ATLAS data \cite{ATLAS:2012cna}, 
$60< \pt^{\gamma} < 90$ GeV, $|\eta^{\gamma}|<1.3$, $\pt^\text{jet}>25$ GeV, $|\eta^\text{jet}|<2.1$, and $\Delta\phi>7\pi/8$. In LBT simulations, we use a Debye screening mass $\mu^2_{D}=4\pi\alpha_s T^2$.


\begin{figure}
\centerline{\includegraphics[width=8.0cm]{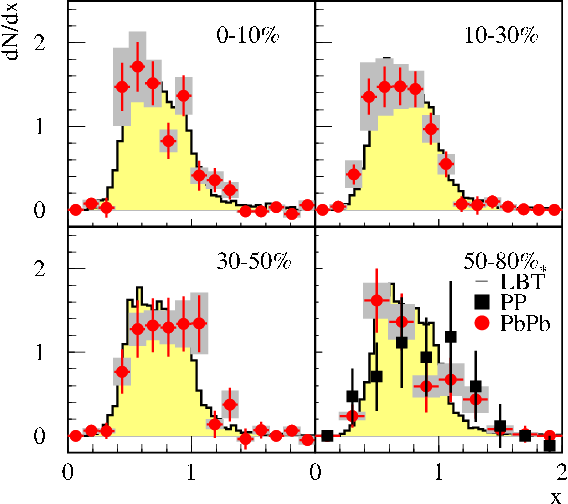}}
 \caption{Distribution of $\gamma$-jet asymmetry $x=\pt^\text{jet}/\pt^\gamma$ in Pb+Pb collisions with four centralities  at $\sqrt{s}=2.76$ TeV from LBT with $\alpha_{ s}=0.2$
  as compared to CMS data \cite{Chatrchyan:2012gt}.
 \label{asym}}
\end{figure}

Following experiments at the LHC \cite{Chatrchyan:2012gt,ATLAS:2012cna}, we first calculate the $\gamma$-jet asymmetry distribution $dN/dx$ with $x=\pt^\text{jet}/\pt^\gamma$. Shown in Fig.~\ref{asym} are $\gamma$-jet asymmetry
distributions from LBT simulations (histogram) as compared to CMS  data \cite{Chatrchyan:2012gt} (solid circles)  in Pb+Pb collisions at $\sqrt{s}=2.76$ TeV with four different centralities and for a jet-cone size $R=0.3$. Because of initial-state radiation, $\gamma$ jets are produced with a large momentum asymmetry in $p+p$ and peripheral Pb+Pb collisions  where there is no or little medium-induced jet energy loss. 
The LBT results can fit the experimental data of both $p+p$ and Pb+Pb with different centralities quite well with a fixed value of $\alpha_{s}=0.2$. 
The asymmetry distributions in $x$ seem to depend very weakly on the centrality even though there is significant jet energy loss as we will show
below.

\begin{figure}
\centerline{\includegraphics[width=8.5cm]{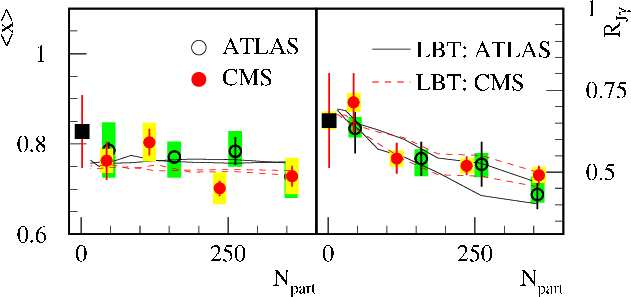}}
 \caption{Averaged $\gamma$-jet asymmetry $\langle x\rangle=\langle \pt^\text{jet}/\pt^\gm\rangle$  (left) and  jet survival rate $R_{J\gamma}$ (right)
 as functions of the number of participant nucleons in Pb+Pb at $\sqrt{s}=2.76$ TeV from LBT as compared to
 experimental data  \cite{Chatrchyan:2012gt,ATLAS:2012cna}. Values of $\alpha_{s}=0.15$--0.23 (dashed line) and 0.2--0.27 (solid line)
 are used for LBT calculations with CMS and ATLAS cuts, respectively.
 \label{avxr}}
\end{figure}

One can further quantify the $\gamma$-jet asymmetry in heavy-ion collisions by the averaged asymmetry or
ratio of the jet and photon transverse momenta $\langle x\rangle=\langle\pt^\text{jet}/\pt^\gm\rangle$ and the jet survival rate or
 fraction of $\gamma$-tagged jets
$R_{J\gamma}$ with $\pt^\text{jet}> 30$ GeV (CMS cut) or $x=\pt^\text{jet}/\pt^\gamma> 0.42$ (ATLAS cut). Shown in Fig.~\ref{avxr}
 are LBT results (lines) on the averaged $\gamma$-jet asymmetry $\left<x\right>$ and jet survival rate $R_{J\gamma}$ as functions of the number of participant nucleons are compared to CMS (solid circles and squares)  \cite{Chatrchyan:2012gt} and ATLAS (open circles) data \cite{ATLAS:2012cna}.
Note those kinematic cuts in ATLAS data, which are also truncated with $x<0.42$, are somewhat different from that in CMS data. 
In LBT calculations, $\alpha_{s}$=0.15--0.23 are used with the CMS cuts (dashed)  while $\alpha_{s}$=0.2--0.27 for the ATLAS cuts (solid). One can see that the averaged momentum asymmetry $\left<x\right>$
has a very weak centrality dependence and is not very sensitive to the value of $\alpha_{s}$. The jet survival rate $R_{J\gamma}$ has, however, a stronger dependence on the centrality and the value of $\alpha_{s}$ or  the strength of the jet-medium
interaction.

\begin{figure}
\centerline{\includegraphics[width=8.9cm]{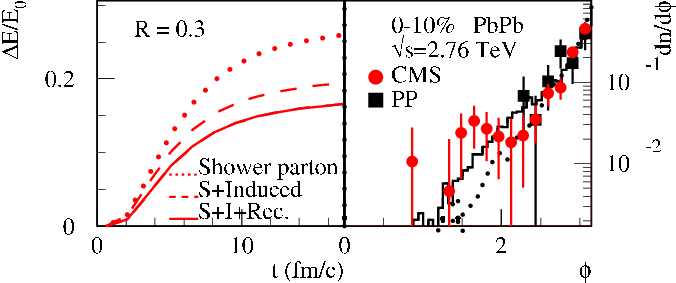}}
 \caption{Averaged energy loss as a function of time (left) and
 azimuthal distribution relative to the $\gamma$ (right) for $\gamma$-tagged jets
 in  central (0\%--10\%) Pb+Pb collisions at $\sqrt{s}=2.76$ TeV.  
  \label{hydro}}
\end{figure}

With the jet-medium interaction strength fixed by the experimental data on $\gamma$-jet asymmetry, we can now turn to the net jet energy loss and azimuthal angle broadening. Shown in Fig.~\ref{hydro} (left) is the averaged jet energy loss  as a function of time in the most central 10\% of Pb+Pb collisions at $\sqrt{s}=2.76$ TeV for a jet-cone size $R=0.3$ from LBT simulations with $\alpha_{s}=0.2$.  Because of rapid cooling due to 3D expansion, jet energy loss saturates approximately 10 fm/$c$ after an initial linear rise with the final fractional jet energy loss of about 15\% (solid line). In such an expanding system, one also expects a reduction in the $p_T$ broadening. As a consequence, the jet azimuthal distribution in central Pb+Pb collisions (solid line) as shown in Fig.~\ref{hydro} (right) remains almost unchanged in the opposite direction of $\gamma$ as compared to that in $p+p$ collisions (dotted line), in agreement with CMS data \cite{Chatrchyan:2012gt} within errors. There is still, however, significant difference between Pb+Pb and $p+p$ collisions at large values of azimuthal angle asymmetry $\Delta\phi=\phi-\pi\equiv \phi_{\rm jet}-\phi_\gamma-\pi$.

Future precision data are therefore necessary to measure the $p_T$ broadening of reconstructed jets. Similar to the results in a uniform medium in Fig.~\ref{uniform}, the inclusion of recoiled partons significantly reduces the net jet energy loss as compared to the case when only shower partons (dotted line) and radiated gluons (dashed line) are included in the jet reconstruction. 

\begin{figure}
\centerline{\includegraphics[width=9cm]{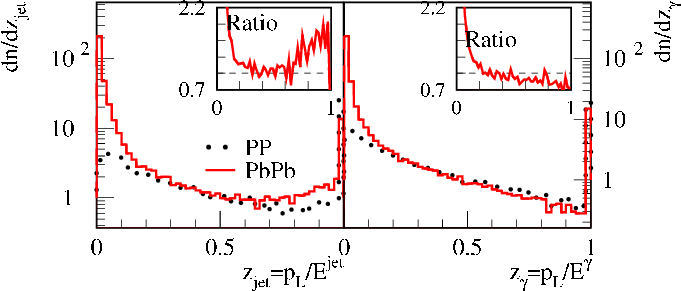}}
 \caption{The reconstructed (left) and $\gamma$-tagged jet fragmentation functions (right)
 within a jet cone $R=0.3$ of $\gamma$-tagged jets in central (0\%--10\%) Pb+Pb collisions at $\sqrt{s}=2.76$ TeV. 
 \label{frag}}
\end{figure}

When shower partons go through multiple scattering and bremsstrahlung, they not only reduce the jet energy
but also change the parton spectrum inside the jet cone. Shown in Fig.~\ref{frag} (left) is what we refer to as the reconstructed jet fragmentation function that uses the reconstructed jet energy to define the momentum fraction $z_{\rm jet}=p_{L}/E_{\rm jet}$.
We can see a large enhancement at small $z_{\rm jet}$ due to soft partons from the bremsstrahlung and medium recoil within the jet-cone. There is almost no medium modification at intermediate $z_{\rm jet}$. Similar features have been seen in experiments 
at the Relativistic Heavy Ion Collider (RHIC) \cite{Putschke:2008wn}. The surprising feature is the enhancement
at large $z_{\rm jet}$ as observed in single jets in central Pb+Pb collisions \cite{CMS:2012wxa}. This feature
is caused by the dominance of leading partons in the reconstructed jet energy, which is used to define the 
momentum fraction $z_{\rm jet}$ \cite{Zapp:2012ak,Majumder:2013re}. 
A large fraction of nonleading partons is transported outside the jet cone leading to
the reduction of the reconstructed jet energy. To verify this explanation, we plot in Fig.~\ref{frag} (right) what we refer to as a
$\gamma$-tagged jet fragmentation function \cite{Wang:1996yh} whose momentum fraction is defined as $z_\gamma=p_{L}/E_\gamma$. 
For fixed value of $E_\gamma$, it more or less reflects the initial jet energy loss before medium modification. We therefore
see significant suppression of the $\gamma$-tagged jet fragmentation function at large $z_\gamma$ as well as 
enhancement at small $z_\gamma$. Such $\gamma$-tagged jet fragmentation function should be more sensitive to the jet-medium interaction and can therefore be used to extract jet transport parameters of the hot medium in heavy-ion collisions. Although our study here is for the parton spectrum within the jet cone, one expects qualitatively the same for hadron distributions.

In summary, we have studied the medium modification of $\gamma$-tagged jets in high-energy heavy-ion collisions within the LBT
model that includes both multiple scattering and medium-induced bremsstrahlung. The model can reproduce well recent experimental data on $\gamma$-jet asymmetry and the survival rate in Pb+Pb collisions at the LHC. Through an examination of jet energy
loss and azimuthal angle broadening of $\gamma$-tagged jets in both a uniform medium and the 3D expanding matter in heavy-ion collisions, we have explained two puzzling features observed in recent experimental data at the LHC. We illustrate within the LBT model that
the rapid cooling of the expanding medium and different length dependences of the jet energy loss and  $p_T$ broadening lead to large $\gamma$-jet asymmetry and yet little azimuthal broadening.  Because some soft partons are transported outside the jet cone causing jet energy loss, the leading partons have increased their contribution to the reconstructed jet energy. This leads to a very small modification to the reconstructed jet fragmentation and even enhancement at large $z_{\rm jet}$. We have further proposed a $\gamma$-tagged jet fragmentation function that uses $E_\gamma$ to define the momentum fraction $z_\gamma$. Its suppression at large $z_\gamma$ is shown to be more sensitive to jet-medium interaction and the jet transport parameter in medium. We also show that the inclusion of recoil partons from jet-induced medium excitation tends to reduce the effective energy loss of a reconstructed jet. Such effects should be included for any precision studies of jet modification.










\begin{acknowledgments}
 We thank M. Cacciari for providing a modified version of {\footnotesize FASTJET} for use in this study.
 This work is supported by the NSFC under Grant No. 11221504, 
U.S. DOE under Contract No. DE-AC02-05CH11231 and within the framework of the JET Collaboration. Y. Z. is also
supported by the German Research Foundation DFG (ITRG) GRK 881 and the Humboldt Foundation.
\end{acknowledgments}


\end{document}